\documentstyle[preprint,aps,psfig]{revtex}

\begin{document}

\title{$Z$-dependent Barriers in Multifragmentation from Poissonian
Reducibility and Thermal Scaling}

\author{L. Beaulieu, L. Phair, L.G. Moretto and G.J. Wozniak}

\address{Nuclear Science Division, Lawrence Berkeley National 
Laboratory, Berkeley, California 94720}

\date{\today}

\maketitle

\begin{abstract}
We explore the natural limit of binomial reducibility in nuclear
multifragmentation by constructing excitation functions for 
intermediate mass fragments (IMF) of a given element $Z$. 
The resulting multiplicity distributions for each window
of transverse energy are Poissonian. Thermal scaling is observed in
the linear Arrhenius plots made from the average 
multiplicity of each element. ``Emission barriers'' are extracted from 
the slopes of the Arrhenius plots and their possible origin is 
discussed.
\end{abstract}


\narrowtext

Emission of multiple intermediate mass fragments (IMF), $3 \le 
Z \le 20$, is an important decay mode in heavy-ion collisions between 
20A and 100A MeV~\cite{Bor92,Mor93}. Despite extensive studies, the 
nature of the fragmentation process, whether statistical or dynamical, 
remains an open problem. An historic overview of low energy reactions 
shows that the emission probabilities and excitation functions are by 
far the best observables in distinguishing between statistical processes 
(dominated by phase space, as in the case of light particle evaporation
and fission), and prompt, dynamical processes (like direct 
reactions)~\cite{Mor69}. Indeed, several aspects of nuclear 
multifragmentation may be understood in terms of 
{\em nearly independent} fragment emission from multifragmenting 
sources with {\em thermal-like} probabilities
~\cite{Moretto95,Tso95,Moretto97,Phair95,Moretto95b,Phair96}.
 
It was found~\cite{Moretto95,Tso95,Moretto97} that the experimental 
$Z$ integrated fragment multiplicity distributions $P_n^m$ are 
binomially distributed in each transverse energy ($E_t$) window, 
where $n$ is the number of emitted fragments and $m$ is the number 
of throws. The transverse energy $E_t$ is calculated from the 
kinetic energies $E_i$ of all the charged particles in an event and 
their polar angles $\theta_i$, as $E_t=\sum _i E_i\sin^2\theta _i$. 
The extracted one-fragment emission probabilities $p$ give linear 
Arrhenius plots (i.e. excitation functions) when $\log 1/p$ 
is plotted vs $1/\sqrt{E_t}$. If the excitation energy $E^*$ is 
proportional to $E_t$ and consequently, the temperature $T$ 
to $\sqrt{E_t}$, these linear Arrhenius plots suggest that $p$ has 
the Boltzmann form $p\propto\exp (-B/T)$~\cite{Moretto95,Tso95,Moretto97}.

Similarly, the charge distributions for each fragment multiplicity $n$ 
are observed to be reducible to a single charge 
distribution and to be thermally scalable~\cite{Phair95,Moretto95b}.
Also, the experimental particle-particle angular correlation is 
reducible to the individual fragment statistical angular 
distributions and thermally scalable~\cite{Phair96}.

The appeal of this comprehensive picture is marred by a number of 
open problems. One problem, which will be dealt with here, is that
the binomial decomposition has been performed on the 
$Z$-integrated fragment multiplicities, typically associated with 
$3 \le Z \le 20$. Thus, the Arrhenius plot generated with the
resulting one fragment probability  $p$ is an average over a range 
of $Z$ values. Fortunately, it has been shown that the Arrhenius plots 
should survive such a $Z$ averaging, and yield an effective ``barrier'' 
(slope) dominated by the lowest $Z$ value~\cite{Moretto95,Tso95,Moretto97}.
However, this procedure clearly implies a substantial loss of 
information, and renders the binomial parameters $p$ and $m$ difficult 
to interpret. 

In light of the above considerations, an analysis of the multiplicities 
for each fragment $Z$ value may solve many of these difficulties. 
Furthermore, it has been pointed out that a binomial distribution could 
be distorted by the averaging associated with the transformation 
$E^* \rightarrow E_t$ leading to possibly incorrect values of $m$ and 
$p$~\cite{Tok97}. However, it has been shown that, while $p$ and $m$ 
separately can conceivably be distorted by the transformation, the 
average multiplicity $\left< n\right> = mp$ is far more resistant to 
the averaging process \cite{Tok97,Phair97}. It would be useful 
if a way could be found of avoiding the individual extraction of $p$ and 
$m$ while retaining the possibility of constructing an Arrhenius plot.
     
In this letter, we analyze the experimental fragment multiplicity 
distributions for each individual fragment $Z$ value. We show that 
they are Poissonian. The associated mean multiplicities for {\em each} 
$Z$ give linear Arrhenius plots from which the corresponding $Z$ 
dependent barriers can be extracted. The physical dependence of these
barriers on $Z$ may shed light on the fundamental physics associated 
with multifragmentation, as fission barriers have done for the fission 
process.

The effect of restricting the fragment definition to a single $Z$ value 
is rather dramatic. In Fig.~\ref{fig1}, ratios of the variance to the
mean as a function of $E_t$ are given for a number of $Z$ values, 
and for the case $Z\ge 3$~\cite{Moretto97}. For individual Z values the 
ratios are very close to one, while for the Z integrated case there is a 
sagging at large $E_t$. The explanation for these features can be found 
by recalling that for a binomial distribution

\begin{equation}
\left< n\right>=mp; \hspace{0.25cm} \sigma ^2=mp(1-p); \hspace{0.25cm}
\frac{\sigma ^2}{\left< n\right>}=1-p.
\end{equation}

For $p$ $\rightarrow$ 0, the ratio $\sigma^2/\left< n\right> \rightarrow$
1. This is the Poisson limit. When extensive summation over $Z$ is 
carried out, the elementary probability $p$ increases sufficiently at 
the highest values of $E_t$ so that the  Poisson distribution is replaced 
by the more general binomial distribution. On the other hand, the 
restriction to any given $Z$ value decreases the elementary probability 
$p$ so dramatically that the above ratio effectively remains unity 
at all values of $E_t$ and the distributions become Poissonian:
\begin{equation}
P_n(Z)=\frac{\left< n\right>^ne^{-\left< n\right>}}{n!},
\end{equation}
where $\left< n\right>$ is $\left< n\right> 
(E_t)$. We show the quality of the Poisson fits to the
multiplicity distribution in Fig.~\ref{fig2}. These Poisson fits 
are excellent for all $Z$ values starting from $Z$=3 up to $Z$=14 
over the entire range of $E_t$ and for all the reactions which we have 
studied. Thus we conclude that reducibility (we should call it now 
Poissonian reducibility) is verified at the level of individual $Z$ 
values for many different systems. Incidentally, for the Xe induced 
reactions, there is an excellent overlap of the data sets for different 
targets as a function of $E_t$. They all follow the Poisson fit to 
the Au target data. The probabilities $P_n$ and the range of $E_t$ 
increase with the increasing target mass from V to Au, as they must 
if $E_t$ is a reasonable measure of the dissipated energy.

The experimental observation of Poissonian reducibility directly 
implies that IMF production is dominated by a stochastic process. 
Of course stochasticity falls directly in the realm of statistical 
decay, either sequential or simultaneous (see section 5.5 of ref.~\cite{Moretto97} or ref.~\cite{Bot95}). It is less clear how 
it would fare within the framework of a dynamical model.

In order to verify thermal scaling i.e. if the emission probabilities 
are thermal, we generate Arrhenius plots by plotting $\log\left< n\right>$ 
vs $1/\sqrt{E_t}$. Here, as in previous works, we assume that $E_t 
\propto E^*$ and that $E^* \propto T^2$, according to the simplest
strongly degenerate Fermi gas dependence at constant volume. We are of 
course aware that high excitation energies and/or lower densities
can lead to deviations, which may well be looked for in the future. 
We expect $\left<n\right>$, like $p$, to be of the form  
$\left<n\right> = F(T,...)e^{-B/T}$, where the specific form of the 
pre-exponential factor depends ultimately on whether a reaction theory 
or a chemical equilibrium description will prevail. We use the Arrhenius 
plot in the traditional spirit of evidentiating the leading $T$ dependence 
contained in the exponential. The top four panels of Fig.~\ref{fig3} gives 
a family of these plots for four different reactions. 
Each family contains $Z$ values extending from $Z$=3 to 
$Z$=14. The observed Arrhenius plots are strikingly linear, and their 
slopes increase smoothly with increasing $Z$ value. One slight 
exception is the large $Z$ ($\ge 10$) data for Xe+Cu. At high $E_t$, 
the data deviates from the linear dependence observed elsewhere. For
this smaller system, it is conceivable that charge conservation 
constraints lead to this behaviour. The overall linear trend demonstrates 
that thermal scaling is also present when individual fragments of a 
specific $Z$ are considered. Even apart from the linearity of the Arrhenius 
plots, important information is already contained in the range covered 
by the yield of individual fragments over the range of $E_t$ shown 
in Fig.~\ref{fig3}. For processes not dominated by phase space 
(e.g. low energy direct reactions), one expects the excitation function 
to depend weakly upon excitation energy. Typically the cross sections
vary by factors of a few. In the present data, the mean multiplicity
$\left<n\right>$ varies with $E_t$ by one to two orders of magnitudes. 
This is a strong evidence for the involvement of the internal degrees 
of freedom typical of high barrier statistical decays. 

The advantage of considering individual Z selected fragments is 
readily apparent. For any given reaction, both Poissonian reducibility 
and thermal scaling are verifiable not just once, as in the binomial 
analysis, but for as many atomic numbers as are experimentally
accessible. Take for example the Ar+Au reaction ($E/A$=110MeV) shown 
in the top right panel of Fig.~\ref{fig3}. For this specific 
reaction, we can verify both reducibility and thermal scaling for 12 
individual atomic numbers. Since there are 29 $E_t$ bins, 
Poissonian reducibility is tested 29 times for each $Z$ value, i.e.,
$12\times 29=348$ times for this reaction alone. Including all the 
cases shown in Fig.~\ref{fig3}, we have tested Poissonian 
reducibility 936 times. This is an extraordinary level of 
verification of the empirical reducibility and thermal scaling with 
the variable  $E_t$. 

Two added bonuses arise from this procedure.

1) The criticism has been raised that the linearity of the
Arrhenius plots arises from an autocorrelation, since the complex 
fragments also contribute to $E_t$ \cite{Tsa97}. In the present 
analysis this criticism can be dismissed, since each individual 
$Z$ contributes a vanishingly small amount to $E_t$ ($\le 5\%$), 
even in the region of maximum yields. Still, to be sure that there 
is no autocorrelation in Fig.~\ref{fig3}, we have repeated the 
analysis, for Xe+Au at 50A MeV, by: i) removing from $E_t$ the 
contribution of the individual $Z$ ($E_t^{Z}$) that we have 
selected (Fig.~\ref{fig3}, bottom left panel); ii) by using only 
the $E_t$ of the light charge particles, $E_t^{LCP}$  (Fig.~\ref{fig3}, 
bottom right panel) . In both cases, the Arrhenius plots remain linear 
over almost the entire range of $E_t$ and cover 1 to 2 orders of 
magnitude. Quantitatively, the rate of change of the slopes with $Z$
remains the same regardless of the definition of $E_t$, as shown in the 
top panel of Fig.~\ref{fig4}. This behaviour is expected if the slopes
are related to some physical barriers.

In our attempt to avoid autocorrelation by excluding from $E_t$ all 
IMFs ($E_t^{LCP}$) or the $Z$ value under investigation ($E_t^Z$), 
we have introduced another kind of distortion. Excluding from 
$E_t$ all fragments of charge $Z$ to produce $E_t^Z$ necessarily 
requires that for those events where $E_t^Z \approx E_t$, the yield 
$n_Z \rightarrow 0$. This produces
the visible turnover of the Arrhenius plots in the bottom panels of 
Fig.~\ref{fig3} (the same argument also applies to $E_t^{LCP}$). It 
has been verified experimentally that the maximum values of the new 
$E_t$ scale do indeed correspond to events in which the contribution 
from a given $Z$ (or all IMFs) is absent.

2) The extracted elementary probability is now $\left< n\right>=
\left<mp\right>$ which, contrary to $p$ and $m$, is very resilient to 
any averaging associated with the transformation from $E^*$ to 
$E_t$~\cite{Tok97,Phair97}.

It may be worth reminding the reader that this procedure does not 
contradict binomial reducibility. To the contrary, it represents 
its natural limit for small values of $p$, and it expands its 
applicability by considering each $Z$ value individually. In going 
from binomial to Poissonian distributions, the price one pays is 
the loss of the parameter $m$. While in many ways this is a convenient 
result, it actually implies a loss of scale. In the time sequential 
interpretation of multifragmentation \cite{Moretto97} this implies 
a loss of information about the time window during which 
multifragmentation occurs in units of the natural channel period, 
or the unit time to which the elementary probability is referred. 
In the spacial interpretation, one loses information about 
the total mass of the source \cite{Moretto97}.

Poissonian reducibility and thermal scaling do not contradict recent
observations regarding the role of reaction dynamics in the {\it{formation}} 
of the hot primary sources~\cite{Monto94,Luka97,Laro97,Tok95,Leco95,Demp96}. 
In particular, the experimental scaling is not affected by the 
presence of multiple sources~\cite{Moretto97} and the analysis 
presented here is a powerful test to establish the degree of 
thermalization in the late stage of the reaction. Kinematic variables 
seem to retain spatial-temporal information about the reaction 
dynamics~\cite{Luka97,Laro97,Tok95,Leco95,Demp96,Bow93} 
while the associated emission probabilities seem to demonstrate, 
as verified nearly a 1000 times in the present work, the role 
of phase space in describing the decay of the sources.

Returning to the Arrhenius plots for individual atomic numbers, it 
is straightforward to obtain the values of the slopes from Fig.~\ref{fig3} 
as a function of $Z$. The interpretation of these slopes as 
``emission barriers'' is very tempting. If we had the correct excitation 
energy, rather than $E_t$, we could obtain the actual barriers as a 
function of $Z$. Unfortunately we are limited to our running variable 
$E_t$, and to the assumption of its proportionality to $E^*$. However, 
the many linear Arrhenius plots shown here cannot be easily explained 
without invoking this proportionality. Therefore, with the necessary
caution, we explore the possible meaning of these ``barriers''. A plot of 
these barriers as a function of $Z$ is potentially rich in information. 
The extracted barriers are shown in Fig.~\ref{fig4} (bottom panel). The 
barriers appear to increase linearly with $Z$ at low $Z$ values and tend 
to sag below linearity at higher ones. 

One could wonder about the role of surface energy on 
the origin of these barriers. Fragments might be thought as forming 
by coalescence into a relatively cold and dense nuclear drop out of 
a hot diluted source. The appearance of a substantial surface energy 
for the fragment would suggest barriers proportional to $Z^{2/3}$ 
($A^{2/3}$). If this were true, then one would expect the barrier for 
each $Z$ to be nearly independent of the system studied. Unfortunately, 
since the relation between $E_t$ and excitation energy is unknown, the 
absolute values of our barriers are also unknown. By normalizing all
systems at $Z$=6 and using the Xe+Au at E/A=50 MeV as the reference, 
one observes barriers that are indeed fairly independent of the 
system (Fig.~\ref{fig4}, bottom panel). Another possibility is to 
compare the dependence of the barriers on $Z$ to that of the conditional 
barriers measured at low energy~\cite{Jing91} (black dots). Their 
similarity with the multifragmentation barriers is dramatically 
illustrated. While the Coulomb-like $Z$ dependence of these barriers 
is suggestive, we should remark that these are emission barriers rather 
than Coulomb barriers. Thus the dominance of the Coulomb term is by 
no means obvious.

In conclusion, Poissonian reducibility and thermal scaling 
of individual fragments of a given $Z$ have been observed 
experimentally for several different systems at bombarding energies
ranging from 50 to 110 MeV/nucleon. The high level of verification 
strongly supports the stochastic/statistical nature of fragment 
production and provides a clear signal for source(s) thermalization 
in the late stage of the reaction. Slope parameters were extracted 
from the Arrhenius plots. The interpretation of these slopes as emission 
barriers, originating either from Coulomb or surface terms, or both, 
still needs to be explored. If the physical significance of these $Z$ 
dependent ``barriers'' must remain lamentably open, there is at least 
the distinct possibility that important physical information is contained 
therein. Data with isotopically resolved light charged particles and IMFs 
are needed to further investigate these phenomena.

Acknowledgements

This work was supported by the Director, Office of Energy Research, 
Office of High Energy and Nuclear Physics, 
Nuclear Physics Division of the US Department of Energy, 
under contract DE-AC03-76SF00098. One of us (L.B)
acknowledge a fellowship from the National Sciences and Engineering
Research Council (NSERC), Canada.

\begin{figure}
\vspace{0.5in}
\caption{The ratio of the variance to the mean number of Li, C, O and 
Ne fragments (solid and open symbols) emitted from the reaction 
$^{36}$Ar+$^{197}$Au at $E/A$=110 MeV. The star symbols show the same 
ratio for all IMFs ($3\le Z\le 20$).}
\label{fig1}
\end{figure}

\begin{figure}
\vspace{0.5in}
\caption{The excitation functions $P_n$ for carbon 
(left column) and neon emission (right column) from the reactions
$^{36}$Ar+$^{197}$Au at $E/A$=110 MeV (top panels) and
$^{129}$Xe+$^{51}$V,$^{\rm nat}$Cu,$^{89}$Y,$^{197}$Au 
(bottom panels). The lines are Poisson fits to the gold target 
data.}
\label{fig2}
\end{figure}

\begin{figure}
\vspace{0.5in}
\caption{Middle and upper panels: The average yield per event of
different elements (symbols) as a function of $1/\protect\sqrt{E_t}$. 
Bottom panels: The Xe+Au data at 50A MeV are replotted using the 
transverse energy of all charged particles excluding the Z that we 
have selected, $E_t^{Z}$ (left), and (right) that only of the light 
charged particles, $E_t^{LCP}$. The lines are fits to the data using 
a Boltzmann form for $\left<n_Z\right>$.}
\label{fig3}
\end{figure}

\begin{figure}
\vspace{0.5in}
\caption{Top panel: Slopes of the Arrhenius plots, normalized to 
$Z=6$, for Xe+Au at 50A MeV as a function of $Z$ using the indicated 
definitions of $E_t$. Bottom panel: The $Z$ dependent ``barriers'' 
(the slopes of the Arrhenius plots in Fig.~\protect\ref{fig3}). The 
``barriers'' have been scaled relative to $Z=6$ of the Xe+Au data. Black 
dots are low energy conditional barriers from ref.~\protect\cite{Jing91} 
normalized to $Z=6$ of Xe+Au.}
\label{fig4}
\end{figure}

\newpage
\begin{figure}
\psfig{figure=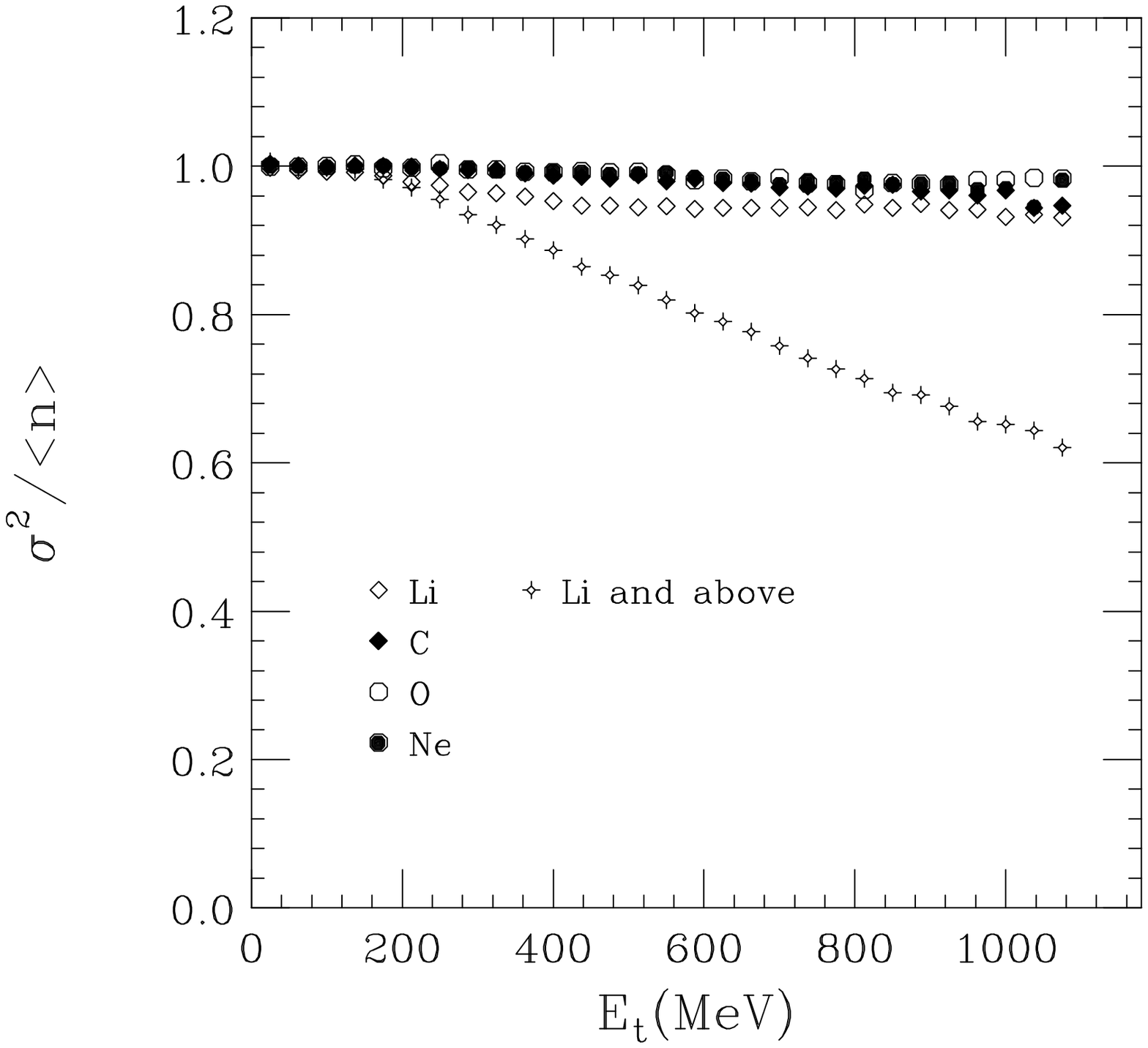,height=6.0in,angle=90}
\vspace{1in}
\center{FIG.\ 1.\ L. Beaulieu {\em et al.}}
\end{figure}

\begin{figure}
\psfig{figure=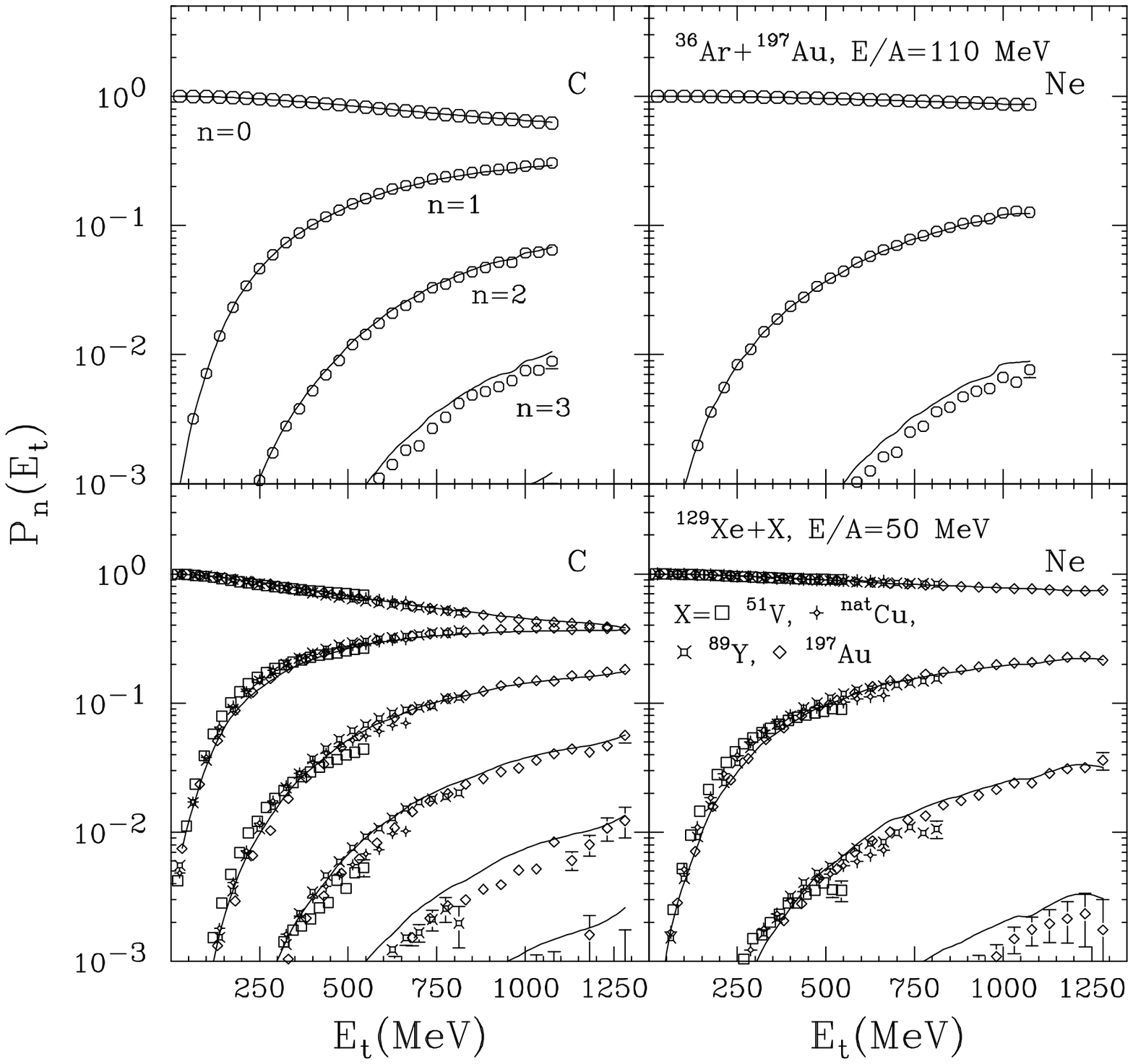,height=6.0in,angle=90}
\vspace{1in}
\center{FIG.\ 2.\ L. Beaulieu {\em et al.}}
\end{figure}

\begin{figure}
\psfig{figure=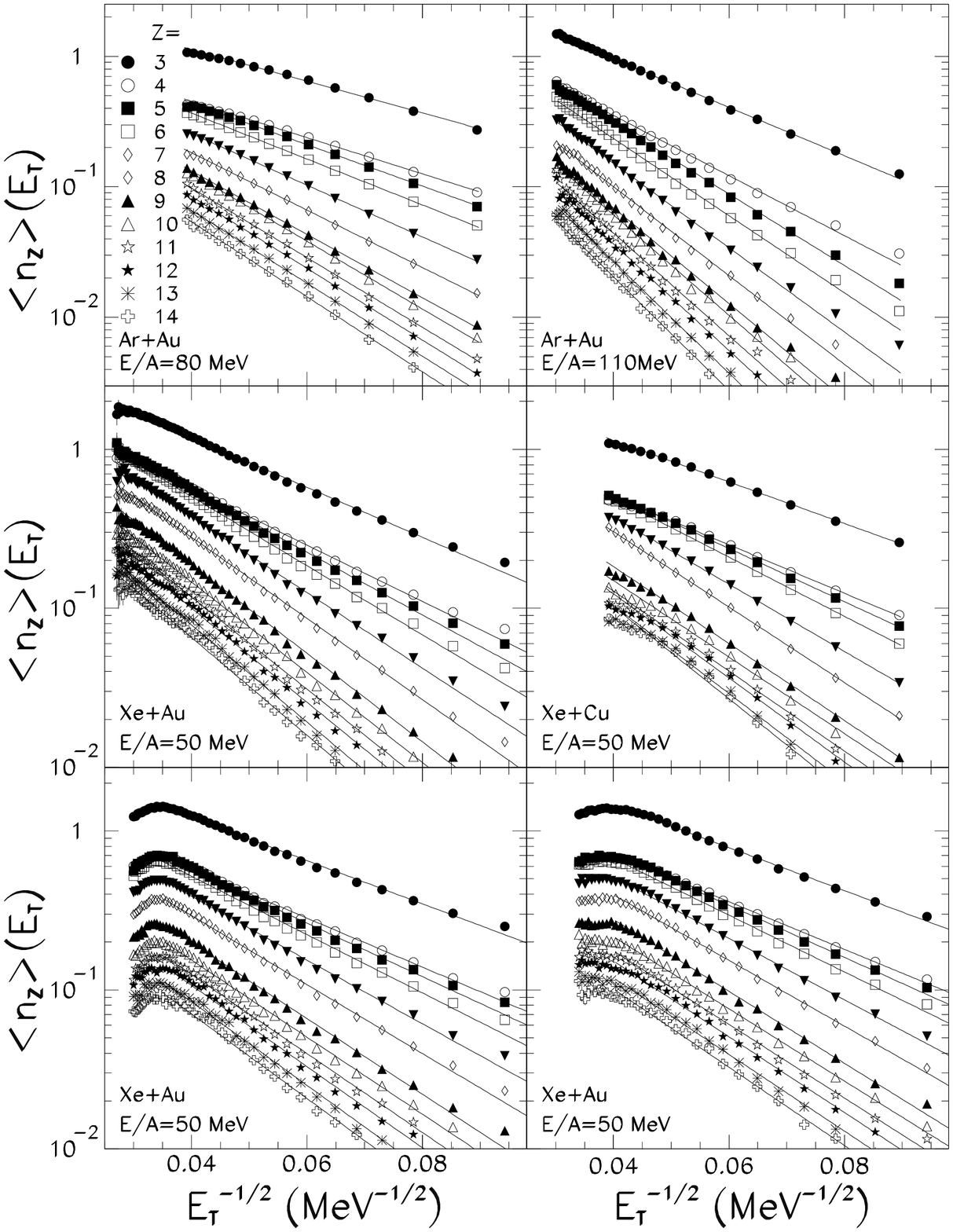,height=7.7in,width=6.0in}
\vspace{1in}
\center{FIG.\ 3.\ L. Beaulieu {\em et al.}}
\end{figure}

\begin{figure}
\psfig{figure=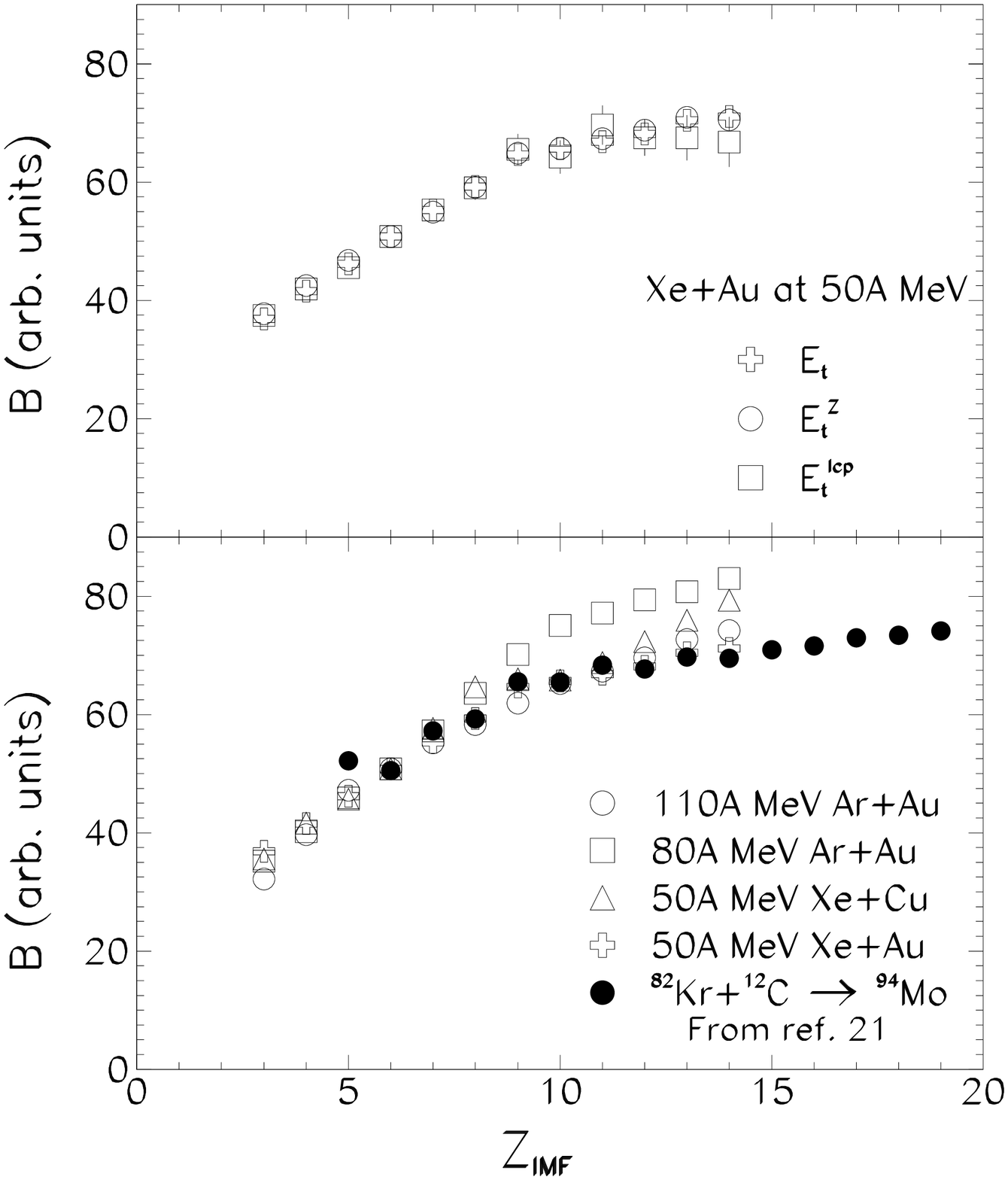,height=7.7in,width=6.0in}
\vspace{1in}
\center{FIG.\ 4.\ L. Beaulieu {\em et al.}}
\end{figure}

\end{document}